\documentclass{article}
\newcommand{\be}{\begin{equation}}
\newcommand{\ee}{\end{equation}}
\newcommand{\ba}{\begin{array}{c}}
\newcommand{\ea}{\end{array}}
\newcommand{\bea}{\begin{eqnarray}}
\newcommand{\eea}{\end{eqnarray}}

\begin{document}

\date{\today}
\title{Distillable entanglement in $d\otimes d$ dimension}
\maketitle
\begin{center}
S. Hamieh\\[0.3cm]
{\it Kernfysisch Versneller Instituut,\\
Zernikelaan
25                                                                              
9747 AA Groningen, 
The Netherlands.}
\end{center}
\vspace*{0.1cm}
\begin{center}
H. Zaraket \\[0.3cm]
{\it Physics Dept, The University of Winnipeg,\\
515 Portage Avenue,
Winnipeg, Manitoba R3B 2E9, Canada.}
\end{center}

\begin{abstract}
Distillable entanglement ($E_d$) is one of the acceptable measures of 
entanglement of mixed states.  
Based on discrimination through local operation and 
classical communication, this paper gives $E_d$ for two classes of orthogonal 
multipartite maximally entangled states. 
\end{abstract}

\section{Introduction}

 Entanglement is believed to be a genuine  resource for quantum computation 
and teleportation. Entanglement shows up in composite quantum systems where 
subsystems do not have {\it pure} states of their own. This is a strict 
quantum phenomena with no classical analogue. However it is highly nontrivial 
to keep such entangled states away from the effect of the environment. 
The environment destroys ``part'' of such entanglement. The residual 
entanglement could be purified or distilled to give maximally entangled 
states. The maximally entangled states (MES), among which there is singlet 
states, can be taken as the basic unit to quantifying entanglement. The 
distillation process uses only local operation ({\it e.g.} unitary 
transformation) and classical communication ({\it e.g.} phone calls) known as 
LOCC. Different LOCC processes could give different number of singlet MES or 
states that could be transformed into singlet state. The maximum number of 
singlet MES is the {\it distillable entanglement}. Distillable 
entanglement is of major 
importance in quantum information processing, the transmission of a pure state
through a noisy channel introduces noise \cite{Benn96} to the initial pure
state and the state becomes mixed. Therefore 
it is important to know the distillable entanglement of the state after  
transmission.\\
For pure bipartite states there is a  
widely acceptable measure of entanglement, the Von Neumann entropy. Whereas 
for mixed states{\footnote{A mixed state is entangled if it cannot be 
represented as a mixture of unentangled pure states.}} different quantities 
are proposed as a measure of the degree of entanglement. Few of these 
quantities survive nowaday, among the acceptable degrees of entanglement is 
the distillable entanglement $E_d$ and the relative entropy of entanglement 
 $E_r$ \cite{Vedr97}. The later is an upper bound for the former \cite{Rain99,Horo00}. This upper bound will be used to calculate $E_d$. Hence 
our strategy in evaluating the distillable entanglement will be\\
\centerline{Number of singlet MES $\leq$ $E_d$ $\leq$ $E_r$.}\\
To saturate the lower bound we use discrimination of {\it orthogonal} MES by 
LOCC as a distillation procedure. It should be mentioned 
that any orthogonal states $|\psi\rangle$ and $|\phi\rangle$ could be discriminated perfectly 
by {\it global} measurement, since they satisfy the necessary condition for 
discrimination $\langle\phi|\psi\rangle=0$. Walgate et al. \cite{Walg00} have 
shown that two orthogonal {\it pure} states  can always be
discriminated by LOCC without need for global measurements. However two 
orthogonal {\it entangled} states can be discriminated by LOCC if only one 
copy is provided. Recently, Ghosh et al.
\cite{Ghos02} have generalized this result, by using the teleportation 
protocol of \cite{Benn93}, and proved that $d$ pairwise  orthogonal MES in $d\,\otimes\,d$ (spin $(d-1)/2$ particles)
 can always be discriminated  by LOCC with a single copy provided.\\
Following the outlined strategy of saturating the lower and upper bound for 
$E_d$ and based on the above results of discrimination in $d\otimes d$, we will 
consider two classes of 
mixtures of orthogonal MES  in $d\, \otimes\, d$, and evaluate their $E_d$.

\section{Four-party state}
The first class of states we consider is the generalization of the state 
studied in \cite{Smol01}   
\be \label{1} \rho = {1\over d}\sum_{i=1}^d 
|\Psi_i\rangle\langle\Psi_i|_{AB} \otimes |\Psi_i\rangle\langle\Psi_i|_{CD}
\,,\ee
where  the states $|\Psi_i\rangle$ are 
any  $d$ pairwise orthogonal MES
chosen out of the $d^2$ 
 orthogonal MES, $\{|\psi_{nm}\rangle$; $n,m=0,1,\dots,d-1\}$, 
in $d\, \otimes\, d$ defined by \cite{Benn93}
\be|\psi_{nm}\rangle= {1\over \sqrt{d}}\sum_{j=0}^{d-1}
e^{2\pi ijn/d}|j\rangle
\otimes |(j+m)\, {\rm mod \,} d\rangle\,,\ee
where $\{{|j\rangle}$; $j=0,1,\dots,d-1\}$ is the standard orthogonal basis of the 
 $d$-dimensional Hilbert space. $\rho$ could be understood as a four party
state where $A$ and $B$ share one of the $d$ orthonormal MES, but don't know
which one (all terms are equally weighted by $1/d$), and $C$ and $D$ share the same 
MES, also not knowing which state they are sharing. 

As in \cite{Ghos01}, in order to compute the distillable entanglement 
of $\rho$ we will use the relative entropy of entanglement
for an entangled quantum state $\sigma$ defined as \cite{Vedr97}
 \be E_r(\sigma)=\min_{\sigma^*\in {\cal D}} S(\sigma || \sigma^*) \,,\ee
where          
${\cal D}$ is the set of all separable states on the Hilbert space
on which $\sigma$ is defined and  $S(\sigma||\sigma^*)= {\rm Tr}\{\sigma(\log_2
\sigma-\log_2\sigma^*)\}$ is the relative entropy  
of $\sigma$ with respect to $\sigma^*$.

 The relative entropy of entanglement for $\rho$ could be 
found by computing the relative entropy of $\rho$ with respect to the equal 
combination of all the $d^2$ MES: 
\be \rho^{S} = {1\over d^2}\sum_{i=1}^{d^2}|\Psi_i\rangle\langle\Psi_i|_{AB}
 \otimes |\Psi_i\rangle\langle\Psi_i|_{CD}\,.\ee
We will show that  $\rho^{S}$ minimizes 
 $S(\rho||\rho^*)$ over $\rho^*\in {\cal D}$. What is particular about 
$\rho^{S}$ is that it is separable across all two party cuts,
$ AC\, :\, BD$, $ AD\, :\, BC$ and by construction across
$ AB\, :\, CD$ (see \cite{Smol01} for the bipartite case, the generalization 
to multipartite case is straight forward). Hence the relative entropy of 
entanglement for $\rho$ across $ AC\, :\, BD$  cut is
\be E_r(\rho)\leq S(\rho||\rho^{S})=\log_2d.\ee
However the distillable entanglement is bounded above by $E_r(\rho)$
\cite{Rain99,Horo00}, so $ E_d(\rho)\leq \log_2d $.
But it has been shown \cite{Ghos02} that it is
 possible to distinguish between the $d$
orthogonal MES by only LOCC if one copy of the
state is provided. C and D can then know which state they have
by LOCC and hence enabling $A$ and $B$ to know with certainty which Bell 
state they share, which could then be transformed by local operation into 
a singlet state, this corresponds to $\log_2d$ ebit{\footnote{For the  
bipartite case a 1 ebit describes any quantum system which contains 
entanglement equivalent to that of a singlet. However a multipartite 
MES of two subsystems A and B has $d$ equally weighted terms in its Schmidt 
decomposition, giving $\log_2d$ of entanglement.}},
  hence    
the distillable entanglement
of $\rho$  in $AC\, : \, BD$ cut is at least $\log_2d$
($E_d(\rho)\geq \log_2d), $
from the lower and upper bound for $E_d$ we deduce that 
\be E_d(\rho)= \log_2d. \ee
This result is a generalization for the bipartite case 
\be \rho = {1\over 2}\sum_{i=1}^2|\Psi_i\rangle\langle\Psi_i|_{AB}
 \otimes |\Psi_i\rangle\langle\Psi_i|_{CD}\,,\ee
where the state $|\Psi_i\rangle$ is one of the four Bell states
\bea \psi_{00}={1\over \sqrt{2}}(|00\rangle+|11\rangle,\nonumber\eea
\bea \psi_{01}={1\over \sqrt{2}}(|01\rangle+|10\rangle,\nonumber\eea
\bea \psi_{11}={1\over \sqrt{2}}(|01\rangle-|10\rangle,\nonumber\eea
\bea \psi_{10}={1\over \sqrt{2}}(|00\rangle-|11\rangle,\eea
and the distillable entanglement
in this case is $E_d(\rho)= 1$ ebit as found in \cite{Ghos01}.

\section{Multi-copy Bill state}

The second class of state we will study is the multi-copy Bill state 
defined by 
\be \label{2} \rho_n = {1\over d^2}\sum_{i=1}^{d^2}|\Psi_i\rangle^{\otimes n}
\langle\Psi_i|
\,,\ee
where each biparty among the $n$-biparties{\footnote{not to be confused with 
bipartite}} shares one of the $d^2$ orthogonal MES, with an equal probability. To 
compute the distillable entanglement
of $ \rho_n$ we will use the method described in \cite{Chen02}
(for the  $2\,\otimes\,2$ case).\\
For $n=1,2$ the state $\rho_n$ is separable
\cite{Smol01} and hence the distillable entanglement is zero.\\
When $n$ is even: $n=2m$ with $m>1$    
the separable state ${\rho_{2}}^{\otimes m} $ could be used to optimize the relative 
entropy of $\rho_{2m}$. Indeed, 
\be \label{3}
 E_d(\rho_{2m})\leq S(\rho_{2m}||{\rho_{2}}^{\otimes m} )=(2m-2)\log_2d.\ee
However, it has been shown \cite{Ghos02} that two copies are necessary 
and sufficient
to discriminate between  $d^2$ orthogonal MES and hence any two biparties
from the $2m$  biparties  can use their two copies of $|\Psi_i\rangle$
 to determine which state they have initially shared and hence
distill $(2m-2)\log_2d$ ebit between the remaining $2m-2$ biparties. 
So the distillation entanglement is at least $(2m-2)\log_2d$. 
 Combining this with Eq.~(\ref{3}) we get
\be  E_d(\rho_{2m})=(2m-2)\log_2d\,.\ee
For odd $n$: $n=2m+1$, an additional step is needed. We first use the fact that one 
can always decompose 
$\rho_{2m+1}^{\otimes2}$ into two copies of the optimal decomposition of 
$\rho_{2m+1}$, {\it i.e.} $2E_r(\rho_{2m+1})\leq E_r(\rho_{2m+1}^{\otimes2})$. 
Then  the relative entropy of $\rho_{2m+1}^{\otimes2}$ is evaluated with respect to 
${\rho_{2}}^{\otimes 2m+1}$, which is separable: 
\be E_r(\rho_{2m+1}^{\otimes 2})\leq S(\rho_{2m+1}^{\otimes 2}||{\rho_{2}}^{\otimes 2m+1})=(4m-2)\log_2d\,,\ee
hence, the distillable entanglement of $\rho_{2m+1}$ is bounded above as 
\be E_d(\rho_{2m+1})\leq  E_r(\rho_{ 2m+1})\leq  
{1\over 2}S(\rho_{2m+1}^{\otimes 2}||{\rho_{2}}^{\otimes
2m+1})=(2m-1)\log_2d \,.\ee
Again, using two copies, out of the $2m+1$ copies, to distinguish between
the $d^2$ $|\Psi_i\rangle$'s we get the lower bound 
$$E_d(\rho_{2m+1})\geq (2m+1-2)\log_2d=(2m-1)\log_2d.$$  
The lower and upper bound for $E_d$ give 
\be  E_d(\rho_{2m+1})=(2m-1)\log_2d\,.\ee
Here we do not venture to 
conjecture any statement concerning the additivity of the relative entropy of 
entanglement for the multipartite case, this will be investigated further in a 
future work.\\
Finally, combining the even and odd $n$ cases, the distillable entanglement of 
$\rho_n$ 
in $d\,\otimes\,d$ is 
\be  E_d(\rho_{n})=(n-2)\log_2d\,.\ee
This is a generalization of the $d=2$ case, studied in \cite{Chen02}, where 
$E_{d=2}(\rho_{n})=n-2$. Again discrimination is shown to be an optimal 
distillation procedure.

\section{Conclusion}
In this note we have made a generalization to $d\,\otimes\,d$ of the
results found in \cite{Ghos01} and \cite{Chen02} for the
distillable entanglement of the four-party state Eq.~(\ref{1}) and the multi-copy 
Bell state Eq.~(\ref{2}) respectively. Discrimination was used as distillation 
procedure and was shown to be optimal, for the specific studied classes of 
states.\\
A natural extension to our work would be the study of more than $d$ states for the 
first class of states we used. For more than $d$ state mixtures discrimination is no longer 
a good candidate for distillation (a mixture of $d+1$ states can not be discriminated 
with one copy provided \cite{Ghos02}). Another important issue is the distillation 
entanglement of orthogonal partially entangled mixtures, this will be addressed 
a in future work.

\section*{Acknowledgments}
We would like to thank G. Kunstatter for carful reading of this manuscript. The 
work of H.~Z. was supported in part by the Natural Sciences and Engineering 
Research Council of Canada.  


\end{document}